\documentclass[useAMS,usenatbib,usegraphicx]{mn2e}

\usepackage{color}
\title [Two close Large Quasar Groups] {Two close Large Quasar Groups of size
$\sim$ 350~Mpc at {$\bf z \sim 1.2$}}

\author[R.G. Clowes et al.]
{Roger G.~Clowes,$^1$\thanks{E-mail: rgclowes@uclan.ac.uk}
Luis E. Campusano,$^2$
Matthew J. Graham$^3$
and \newauthor
Ilona K. S\"ochting$^4$ \\
$^1$ Jeremiah Horrocks Institute, University of Central Lancashire,
Preston PR1 2HE \\
$^2$ Observatorio Astron\'omico Cerro Cal\'an, Departamento de Astronom\'{\i}a,
Universidad de Chile, Casilla 36-D, Santiago, Chile \\
$^3$ California Institute of Technology, 1200 East California Boulevard,
Pasadena, CA 91125, USA \\
$^4$ Astrophysics, Denys Wilkinson Building, Keble Road,
University of Oxford, Oxford OX1 3RH}
\begin{document}

\date{Accepted 2011 Xxxxxxx XX. Received 2011 Xxxxxxx XX; in original form
2011 Xxxxxxx XX}

\pagerange{\pageref{firstpage}--\pageref{lastpage}} \pubyear{2011}

\maketitle

\label{firstpage}

\begin{abstract}
The \citet{Clowes1991} Large Quasar Group (LQG) at $\bar{z} = 1.28$ has been
re-examined using the quasar data from the DR7QSO catalogue of the Sloan
Digital Sky Survey. In the 1991 discovery, the LQG impinged on the northern,
southern and eastern limits of the survey. In the DR7QSO data, the western,
northern and southern boundaries of the LQG remain essentially the same, but
an extension eastwards of $\sim 2^\circ$ is indicated. In the DR7QSO data,
the LQG has 34 members, with $\bar{z} = 1.28$. A new group of 38 members is
indicated at $\bar{z} = 1.11$ and within $\sim 2.0^\circ$ of the Clowes \&
Campusano LQG. The characteristic sizes of these two LQGs, $\sim$
350-400~Mpc, appear to be only marginally consistent with the scale of
homogeneity in the concordance cosmology. In addition to their intrinsic
interest, these two LQGs provide locations in which to investigate early
large-scale structure in galaxies and to identify high-$z$ clusters. A method
is presented for assessing the statistical significance and overdensity of
groups found by linkage of points.
\end{abstract}


\section{Introduction}

Since 1982 there have been many reports of large-scale structures (LSSs) in
the distribution of quasars and related objects. These structures have now
become generally known as Large Quasar Groups (LQGs). They are the largest
structures so far seen in the early universe ($z \sim$ 0.4--2.0), with sizes
in the range 70--250~Mpc, and memberships $\ga 5$. The \citet{Clowes1991}
LQG, at redshift $z \sim 1.3$ and with a longest dimension of $\sim$ 250~Mpc,
is a particularly large example of LSS in the early universe. For some of the
historical development of work on LQGs see, for example: \citet{Webster1982};
\citet*{Crampton1987}, \citet*{Crampton1989}; \citet{Clowes1991};
\citet{Komberg1994}; \citet*{Graham1995}; \citet*{Komberg1996};
\citet{Newman1998}, \citet{Newman1999}; \citet*{Clowes1999};
\citet{Tesch2000}; \citet{Williger2002}; \citet*{Haines2004};
\citet{Brand2003} (for radio galaxies); \citet{Miller2004}; and
\citet{Pilipenko2007}.

LQGs are of interest not only as examples of large-scale features in the
early universe but also because of their potential for: investigations of the
LSS in galaxies; for indicating places in which to look for high-$z$
clusters; and for identifying the environments that favour the formation of
quasars.

\citet{Komberg1996} \citep[see also][]{Komberg1994} considered that LQGs
denote the precursors at high redshifts of the superclusters seen today.
\citet{Pilipenko2007} similarly considered that LQGs may be ``incipient
superclusters'' and also concluded that a substantial fraction of quasars lie
two-dimensionally, in sheets. Among some of the earliest papers in this area,
\citet{de-Ruiter1982} considered the possibility that quasars reside in
superclusters to be a natural explanation for the occurrence of wide-angle
doublets and triplets of quasars of very similar redshifts. \citet{Longo1991}
showed that the ``Great Wall'' of galaxies \citep{Geller1989} at $z \sim
0.03$ is traced by its AGN. More recently, \citet{Mountrichas2009} found that
quasars and luminous red galaxies cross-correlate on scales $\la$ 40~Mpc for
$0.35 \le z \le 0.75$. \citet*{Soechting2002} and \citet*{Soechting2004}
showed that, at the low redshifts $z \sim 0.3$, quasars tended to follow the
LSS in clusters of galaxies, but were preferentially associated with the
peripheries of clusters. \citep[See also][and further references in that
  paper.]{Sanchez1999} For the \citet{Crampton1989} LQG at $z \sim 1.1$, a
similar result was found by \citet{Tanaka2001} for five of its quasars: they
followed a chain of clusters (or groups) but were associated with the
peripheries. \citet{Ohta2003} found three quasars in the vicinity of a
supercluster at $z = 1.27$ that are presumed to be associated, while being
separated from the two component clusters by $\sim$ 7--17 Mpc.

For the Clowes \& Campusano LQG, \citet{Haines2001} found that one quasar is
simultaneously on the peripheries of two clusters (or groups) of red
galaxies, together with a band of blue, presumably star-forming, galaxies. The
association of two further quasars from the Clowes \& Campusano LQG with
clusters or their peripheries is less apparent \citep{Haines2004}, and one at
least of these may be quite isolated. There is evidence for general
sheet-like enhancements of red galaxies with embedded clusters associated
with this LQG \citep{Haines2004}. \citet{Haberzettl2009} similarly find
evidence for enhancements of candidates for Lyman-break galaxies associated
with it. A direct association (i.e.\ using spectroscopic redshifts rather
than photometric estimates) of the Clowes \& Campusano LQG with an excess of
galaxies has been achieved via MgII absorbers in background quasars, which
showed a $\sim$ 200 per cent excess in the redshift interval $1.2 < z < 1.4$
\citep{Williger2002}.

The investigation of LQGs was for many years made difficult by the
limitations of quasar surveys. Surveys tended to be small, or larger but not
very deep, or affected by selection effects, and so on. Compilations of
different surveys increased the numbers of quasars but generally worsened the
selection effects. In recent years, however, the difficulties with quasar
data have been substantially lessened by the Sloan Digital Sky Survey
\citep[SDSS, e.g.][]{Schneider2010, Schneider2007, Richards2006,
  Vanden-Berk2005} and the Two-Degree Field (2dF) QSO Redshift Survey
\citep[2QZ, e.g.][]{Croom2004, Smith2005}.

\citet{Miller2004} and \citet{Pilipenko2007} have investigated LSS in the 2QZ
quasars, with Miller et al.\ finding $\sim$ 200~Mpc structures in a
statistical sense, and Pilipenko more specifically giving coordinates and
other properties of the LQGs discovered. In particular, Pilipenko suggests
that there might be two categories of LQG: (i) size $\sim$ 85~Mpc, membership
$\sim$ 6-8, overdensity $\sim$ 10; and (ii) size $\sim$ 200~Mpc, membership
$\ga 15$, overdensity $\sim$ 4. Note that these overdensities are
substantially larger than those found by \citet{Miller2004} using spherical
filtering.

The \citet{Clowes1991} LQG is, as mentioned above, a particularly large
structure in the early universe. It was discovered by effectively random
spatial sampling of quasar candidates from an objective-prism survey. This
sampling was necessary for homogeneous coverage of a large area ($\sim$ 25
deg$^2$) by single-object spectroscopy at a time when it was not feasible
with multi-object spectroscopy.

The field of the Clowes \& Campusano LQG is now contained within the SDSS.
This paper considers what additional properties of the LQG and its
cosmological neighbourhood can be deduced using the SDSS quasars, given
their wide-angle coverage and almost complete multi-object spectroscopy.

The results presented here indicate that both the Clowes \& Campusano LQG and
a newly-discovered neighbouring LQG are among the largest features so far
seen in the early universe. In this context, we note that there have been
some reports of still larger correlations. \citet{Nabokov2008} presented
preliminary evidence for Gpc-scale correlations of galaxies, and
\citet{Padmanabhan2007} and \citet*{Thomas2011} both found power on
Gpc-scales in the power spectrum of SDSS galaxies. A particularly striking
result is that of \citet{Hutsemekers2005}, who found that the polarisation
vectors of quasars are correlated on Gpc-scales.

Note that the concordance model is adopted for cosmological calculations,
with $\Omega_T = 1$, $\Omega_M = 0.27$, $\Omega_\Lambda = 0.73$, and $H_0 =
70$~kms$^{-1}$Mpc$^{-1}$. All sizes given are proper sizes at the present
epoch.

\section{The SDSS quasars}

The SDSS DR7 quasar catalogue \citep[``DR7QSO'',][]{Schneider2010} of 105783
quasars has been used for this work. Schneider et al. note that the catalogue
does not constitute a statistical sample (i.e.\ a sample with homogeneous
selection) and refer to \citet{Vanden-Berk2005} and \citet{Richards2006} for
detailed discussion of the important properties of completeness and
efficiency of the parent survey. In particular, \citet{Richards2006} describe
how to construct a statistical sample from the DR3QSO catalogue, in a
discussion which should also be applicable to the DR7QSO catalogue. However,
because the requirement here is for assessing the spatial connectivity of
quasars across intervals on the sky of $\sim 5^\circ$ and intervals of
redshift $\Delta z \sim 0.2$, rather than for the luminosity function, the
criteria of Richards et al.\ can be relaxed. Firstly, the redshifts $z \la 2$
of main interest are well within the limit $z \la 3$ of the low-redshift
strand of selection so the changes in the SDSS selection algorithms from the
initial to the final \citep{Richards2002} should not be important. Secondly,
satisfactory spatial uniformity of selection on the sky for these redshifts
should be achievable by selecting $i \le 19.1$ \citep{Richards2006,
  Vanden-Berk2005}, since those quasars are predominantly from the
low-redshift selection strand, which is limited to $i \le 19.1$.

For this re-examination of the Clowes \& Campusano LQG we use the entire
DR7QSO catalogue of $\sim$ 9380~deg$^2$, limited to $i \le 19.1$. The
coverage of the catalogue comprises a large main area for the north galactic
cap (NGC, $\sim$ 7600~deg$^2$) with some jagged boundaries, three equatorial
stripes (totalling $\sim$ 800~deg$^2$) and several distinct, smaller areas
(``special plates''). The LQG is in the main NGC area, and is $\sim$ 5~deg
from the nearest boundary. Some nearby holes in the coverage that could be
seen in the preceding DR5QSO catalogue appear to be no longer present in
DR7QSO.  Of course, any candidate LQG elsewhere in the catalogue that
encounters a boundary might not be completely identified.

We shall denote the area of $\sim$ 9380~deg$^2$ as A9380. We also define a
control area, designated A3725, of $\sim$ 3725 deg$^2$ (actually 3724.5
deg$^2$) by RA: $123.0^\circ \rightarrow 237.0^\circ$ and Dec: $15.0^\circ
\rightarrow 56.0^\circ$. A3725 is chosen to be a large area well separated
from the LQG.

Note that the relatively bright limiting magnitude, $i \le 19.1$, of the SDSS
low-redshift selection strand is not ideal for the tasks of finding and
further investigating LQGs. The difficulty is possibly illustrated by
\citet{Pilipenko2007}, who finds LQGs in the fainter 2QZ data \citep[see
  also][]{Miller2004} but finds no significant groups beyond doublets and
triplets in the DR5QSO \citep{Schneider2007}, although the adoption of a
small linkage scale (57~Mpc compared with mean separation of 83~Mpc) for the
DR5QSO seems likely to have also been a very substantial factor in their
apparent absence. The identification of LSSs by algorithms can be quite
subtle, especially concerning the effective and objective specification of
factors such as the linkage scale and overdensity.

The discovery of the Clowes \& Campusano LQG was reported in
\citet{Clowes1991}, with a later revision, following some additional
observations, given by \citet{Clowes1999}. From these two papers, the LQG was
detected as 18 quasars in the redshift range $1.2 \le z < 1.4$, of which 15
were new discoveries and three were previously known. At the time of the
observations multi-object spectroscopy across the whole survey area of 25.3
deg$^2$ was not practicable and, instead, wide-area coverage was achieved by
effectively random spatial sampling of the quasar candidates for
single-object spectroscopy. Although the magnitude limit of $i = 19.1$ for
the SDSS low-redshift strand is brighter than the Clowes \& Campusano limit
of $B_J \sim 20.4$ there should be many further SDSS quasars in the range
$1.2 \le z \le 1.4$ because of the complete wide-area coverage. Note that of
the 18 original LQG members, 10 are present in the DR7QSO catalogue with $i
\le 19.1$, and a further two are present with $i > 19.1$.

Ten further quasars with $1.2 \le z < 1.4$ in the region of the LQG are known
from a later UV-excess survey \citep[the Chile-UK Quasar Survey:][]{Newman1998,
  Newman1999} with $b_J \le 20$ and of incomplete spatial coverage. Of these
10, eight are present in the DR7QSO catalogue, all with $i \le 19.1$. (One of
these eight has a DR7QSO redshift slightly greater than 1.4.)

In \citet{Clowes1991} the assessment of significance of the LQG made use of
the two-dimensional minimal spanning tree (MST) of the RA, Dec coordinates.
This paper uses three-dimensional single-linkage hierarchical clustering,
which is equivalent to the three-dimensional MST. These MST-type algorithms
have the great advantage that they do not require assumptions about the
morphology of the structure. With their use there are at least three
parameters needed to specify the LQGs --- the linkage scale, the number of
members, and the overdensity. Usually the analyses concentrate on a chosen
linkage scale. \citet{Pilipenko2007} discusses two possibilities for making
the choice: (i) the physical, in which one chooses the scale that maximises
the fraction of groups that match closely some specified physical parameters
(e.g.\ size, membership, density); and (ii) the formal, in which, for
example, one might choose the scale that maximises the number of groups found
\citep[e.g.][]{Graham1995}. Note that some LQGs found at one linkage scale
may be fragments of LQGs that would be found completely at a larger linkage
scale.

The mean nearest-neighbour separation is an objective measure that can be
used to guide the choice of linkage scale. For the redshifts of interest
here, $z \sim 1.3$, the mean nearest-neighbour separation calculated from
area A3725 (as $0.55\rho^{-1/3}$) is $\sim$ 74~Mpc. Note that this
nearest-neighbour separation is a true, physical, separation, determined
globally by the number of points in a particular volume. However, the linkage
in practice between any two quasars is an apparent separation incorporating
the true separation plus distortions that arise from the contributions to the
redshifts of observational uncertainties and peculiar velocities. Several
papers quote estimates for the observational redshift uncertainties of SDSS
quasars: 0.004 \citep[][DR7]{Schneider2010}; 0.006
\citep[][DR5]{Schneider2007}; $< 0.01$ \citep[][DR3, $z \la
  3.4$]{Trammell2007}; $\sim$ 0.01 \citep[][DR5, $z \ge 2.9$]{Shen2007};
0.003--0.01 \citep[][compilation]{Ross2009}. The value of 0.006 from
\citet{Schneider2007} refers to a difference of redshifts so we should divide
it by $\surd 2$ to obtain the uncertainty for a single measurement, 0.004,
which is the same as the value from \citet{Schneider2010}. We adopt the
\citet{Schneider2010} uncertainty of $\Delta z_{obs} \sim 0.004$ as a
representative DR7 value. The mean redshift of the DR7QSO quasars with $i \le
19.1$ is $z = 1.38$ so this value of the uncertainty should be appropriate to
the redshifts of interest here. Note that the SDSS data processing
\citep{Schneider2007, Stoughton2002} attempts to correct for the redshift
offsets between high- and low-ionisation emission lines \citep[$\sim$
  600~kms$^{-1}$,][]{Gaskell1982} and so this possible source of distortion
is not considered further here. \citet{Hewett2010} have provided a catalogue
of revised redshifts for SDSS DR6 quasars that corrects for small systematic
offsets in the SDSS redshifts, reduces the redshift uncertainties, and
provides an estimate of the redshift uncertainty for the individual quasars
(typically $\sim 0.0006$). However, DR6 contains large gaps in the sky
coverage compared with DR7.

If we assume that quasars have peculiar velocities in the radial direction of
$\sim$ 400~kms$^{-1}$ then there is a further random component to the
redshift $\Delta z_{pec} \sim 0.003$ at $z \sim 1.3$\footnote{This estimate,
  provided by O.~Snaith (private communication) from simulations, refers to
  galaxies on the outskirts of clusters (as indicated for quasar
  environments) at $z \sim 1.3$.}. Combining $\Delta z_{obs} \sim 0.004$ and
$\Delta z_{pec} \sim 0.003$ in quadrature, gives the typical distortion in
position of a quasar of $\sim$ 11~Mpc at $z \sim 1.3$. The corresponding
distortion in pairwise separations is then $\sim$ 16~Mpc. Thus a set of
quasars that forms a unit at a true linkage scale of 73~Mpc would be very
likely to appear fragmented for apparent linkage scales $\sim$ 90~Mpc.

The detection of the LQG in the DR7QSO catalogue is considered below for
linkage scales in the range 75--105~Mpc. The upper limit of this range is set
to be smaller than the expected percolation radius for a Poisson distribution
\citep{Martinez2002, Pike1974}, which, for $z \sim 1.3$ and A3725, is $\sim$
115~Mpc.  This range of 30~Mpc was explored by a binary chop rather than by a
series of equal increments. The redshift distribution for the DR7QSO
catalogue with $i \le 19.1$ is fairly flat across the interval 1.0--1.8 and
so this is the range that has been considered in practice, rather than simply
1.2--1.4.

Using the {\it agnes\/} algorithm in the {\it R package}\footnote{See
  http://www.r-project.org} for single-linkage hierarchical clustering, a
unit of 34 quasars emerges at a linkage scale of 100~Mpc that, given evidence
presented below, appears to be the LQG. At the other linkage scale considered
in the binary chop, 90~Mpc, the LQG does not appear at the specified minimum
cluster size of 10 members. The apparent linkage scale of 100~Mpc thus means
that the LQG is unlikely to be fragmenting at the true mean nearest-neighbour
separation $\sim$ 73~Mpc. (Note that this scale of 100~Mpc is between the two
scales \{$100h^{-1}$, $200h^{-1}$~Mpc diameters, equivalent to $71$, $143$~Mpc
radii\} for spherical filtering used by \citealt{Miller2004} to find LQGs.)

The 34 quasars connected at this 100~Mpc linkage scale are listed in
Table~\ref{lqg_U1.28_table}. Their mean redshift is 1.28, identical to that
of the original 18 members of the LQG. They cover the redshift range $1.1865
\rightarrow 1.4232$, whereas the original 18 cover the range $1.207
\rightarrow 1.386$.

The centroids of this unit of 34 and of the original 18 are separated by only
$\sim 0.92^\circ$.  The identification as the LQG of this unit of 34,
occurring at the correct redshift, within $0.92^\circ$, and also (see the
next section) with essentially the same western, northern and southern
boundaries, therefore seems certain.

For convenience in what follows, this LQG unit of 34 quasars will be
designated U1.28 from its mean redshift, and similarly for two other units of
interest that are discussed below.

\begin {table*}
\flushleft
\caption {U1.28: the set of 34 100Mpc-linked quasars from the SDSS DR7QSO
catalogue that are associated with the Clowes \& Campusano (1991) LQG. The
columns are: SDSS name; RA, Dec. (2000); redshift; $i$ magnitude; comments.}
\small
\renewcommand \arraystretch {0.8}
\newdimen\padwidth
\setbox0=\hbox{\rm0}
\padwidth=0.3\wd0
\catcode`|=\active
\def|{\kern\padwidth}
\newdimen\digitwidth
\setbox0=\hbox{\rm0}
\digitwidth=0.7\wd0
\catcode`!=\active
\def!{\kern\digitwidth}
\begin {tabular} {lllll}
\\
SDSS name            & RA, Dec (2000)             & z      & $i $   & Comments \\
                     &                            &        &        &          \\
\\
103744.89$+$051834.2 & 10:37:44.89~~$+$05:18:34.2 & 1.2280 & 18.958 & b        \\
104114.06$+$034312.0 & 10:41:14.06~~$+$03:43:12.0 & 1.2633 & 18.588 & b        \\
104115.58$+$051345.0 & 10:41:15.58~~$+$05:13:45.0 & 1.2553 & 18.697 & a        \\
104116.79$+$035511.4 & 10:41:16.79~~$+$03:55:11.4 & 1.2444 & 18.531 & b        \\
104149.92$+$064336.5 & 10:41:49.92~~$+$06:43:36.5 & 1.3238 & 18.923 & a        \\
104225.63$+$035539.1 & 10:42:25.63~~$+$03:55:39.1 & 1.2293 & 17.879 & b        \\
104256.38$+$054937.4 & 10:42:56.38~~$+$05:49:37.4 & 1.3555 & 18.661 &          \\
104304.95$+$052515.6 & 10:43:04.95~~$+$05:25:15.6 & 1.1865 & 18.908 &          \\
104321.88$+$045920.6 & 10:43:21.88~~$+$04:59:20.6 & 1.3646 & 19.001 &          \\
104345.39$+$040300.3 & 10:43:45.39~~$+$04:03:00.3 & 1.1884 & 18.937 &          \\
104425.80$+$060925.6 & 10:44:25.80~~$+$06:09:25.6 & 1.2523 & 18.652 & b        \\
104426.79$+$072754.9 & 10:44:26.79~~$+$07:27:54.9 & 1.4232 & 18.846 &          \\
104445.32$+$054348.8 & 10:44:45.32~~$+$05:43:48.8 & 1.1879 & 18.793 &          \\
104556.93$+$072714.7 & 10:45:56.93~~$+$07:27:14.7 & 1.3966 & 18.907 &          \\
104637.30$+$075318.7 & 10:46:37.30~~$+$07:53:18.7 & 1.3635 & 17.612 &          \\
104656.71$+$054150.3 & 10:46:56.71~~$+$05:41:50.3 & 1.2284 & 17.594 & a        \\
104733.16$+$052454.9 & 10:47:33.16~~$+$05:24:54.9 & 1.3341 & 17.705 & a        \\
104752.69$+$061828.9 & 10:47:52.69~~$+$06:18:28.9 & 1.3125 & 18.954 & a        \\
104843.05$+$064456.8 & 10:48:43.05~~$+$06:44:56.8 & 1.3523 & 18.721 &          \\
105010.05$+$043249.1 & 10:50:10.05~~$+$04:32:49.1 & 1.2158 & 18.151 & a        \\
105018.10$+$052826.4 & 10:50:18.10~~$+$05:28:26.4 & 1.3067 & 19.074 & b        \\
105022.81$+$064621.8 & 10:50:22.81~~$+$06:46:21.8 & 1.2900 & 18.362 & b        \\
105149.58$+$033430.2 & 10:51:49.58~~$+$03:34:30.2 & 1.2697 & 19.044 &          \\
105422.47$+$033719.3 & 10:54:22.47~~$+$03:37:19.3 & 1.2278 & 17.972 &          \\
105423.26$+$051909.8 & 10:54:23.26~~$+$05:19:09.8 & 1.2785 & 18.283 &          \\
105512.23$+$061243.9 & 10:55:12.23~~$+$06:12:43.9 & 1.3018 & 18.413 &          \\
105534.66$+$033028.8 & 10:55:34.66~~$+$03:30:28.8 & 1.2495 & 18.195 &          \\
105537.63$+$040520.0 & 10:55:37.63~~$+$04:05:20.0 & 1.2619 & 18.651 &          \\
105719.23$+$045548.2 & 10:57:19.23~~$+$04:55:48.2 & 1.3355 & 18.429 &          \\
105810.30$+$025145.7 & 10:58:10.30~~$+$02:51:45.7 & 1.2761 & 18.842 &          \\
105821.28$+$053448.9 & 10:58:21.28~~$+$05:34:48.9 & 1.2540 & 18.134 &          \\ 
105833.86$+$055440.2 & 10:58:33.86~~$+$05:54:40.2 & 1.3222 & 18.758 &          \\
110108.00$+$043849.6 & 11:01:08.00~~$+$04:38:49.6 & 1.2516 & 18.254 &          \\
110412.00$+$044058.2 & 11:04:12.00~~$+$04:40:58.2 & 1.2554 & 18.851 &          \\
\\
\end {tabular}
\\
a $\rightarrow$ Members of the original LQG. See \citet{Clowes1991},
\citet{Clowes1994}, and \citet{Clowes1999}.                                \\
b $\rightarrow$ Possible additional members of the original LQG from a survey
with incomplete spatial coverage: \citet{Newman1998} and
\citet{Newman1999}.                                                        \\
\label{lqg_U1.28_table}
\end {table*}

\begin {table*}
\flushleft
\caption {U1.11: the set of 38 100Mpc-linked quasars from the SDSS DR7QSO
catalogue.  The columns are: SDSS name; RA, Dec. (2000); redshift; $i$
magnitude.}
\small
\renewcommand \arraystretch {0.8}
\newdimen\padwidth
\setbox0=\hbox{\rm0}
\padwidth=0.3\wd0
\catcode`|=\active
\def|{\kern\padwidth}
\newdimen\digitwidth
\setbox0=\hbox{\rm0}
\digitwidth=0.7\wd0
\catcode`!=\active
\def!{\kern\digitwidth}
\begin {tabular} {lllll}
\\
SDSS name            & RA, Dec (2000)             & z      & $i $              \\
                     &                            &        &                   \\
\\
102907.22$+$021552.4 & 10:29:07.22~~$+$02:15:52.4 & 1.0173 & 19.050            \\
103200.20$+$022056.4 & 10:32:00.20~~$+$02:20:56.4 & 1.0038 & 18.901            \\
103300.11$+$042116.9 & 10:33:00.11~~$+$04:21:16.9 & 1.0144 & 17.875            \\
103552.43$+$032537.2 & 10:35:52.43~~$+$03:25:37.2 & 1.0553 & 18.980            \\
103626.33$+$045436.4 & 10:36:26.33~~$+$04:54:36.4 & 1.0477 & 18.404            \\
103639.63$+$022553.5 & 10:36:39.63~~$+$02:25:53.5 & 1.0525 & 18.817            \\
103641.96$+$050941.1 & 10:36:41.96~~$+$05:09:41.1 & 1.0631 & 18.553            \\
103709.33$+$022055.7 & 10:37:09.33~~$+$02:20:55.7 & 1.0143 & 18.449            \\
103739.49$+$034946.9 & 10:37:39.49~~$+$03:49:46.9 & 1.0321 & 18.924            \\
103743.97$+$040233.8 & 10:37:43.97~~$+$04:02:33.8 & 1.0932 & 18.671            \\
103748.36$+$040242.1 & 10:37:48.36~~$+$04:02:42.1 & 1.0869 & 17.857            \\
103806.57$+$020234.3 & 10:38:06.57~~$+$02:02:34.3 & 1.0526 & 19.068            \\
104012.14$+$043904.6 & 10:40:12.14~~$+$04:39:04.6 & 1.1195 & 18.578            \\
104309.70$+$075317.8 & 10:43:09.70~~$+$07:53:17.8 & 1.1823 & 18.872            \\
104410.13$+$072305.6 & 10:44:10.13~~$+$07:23:05.6 & 1.1514 & 18.189            \\
104446.16$+$070651.4 & 10:44:46.16~~$+$07:06:51.4 & 1.1292 & 17.973            \\
104506.44$+$051627.4 & 10:45:06.44~~$+$05:16:27.4 & 1.1116 & 18.753            \\
104509.93$+$063559.0 & 10:45:09.93~~$+$06:35:59.0 & 1.1184 & 19.001            \\
104636.93$+$082437.4 & 10:46:36.93~~$+$08:24:37.4 & 1.1398 & 18.747            \\
104752.93$+$022408.1 & 10:47:52.93~~$+$02:24:08.1 & 1.1390 & 19.012            \\ 
104835.72$+$000002.3 & 10:48:35.72~~$+$00:00:02.3 & 1.1429 & 18.641            \\
104901.71$+$005534.0 & 10:49:01.71~~$+$00:55:34.0 & 1.1630 & 18.215            \\
104932.22$+$050531.7 & 10:49:32.22~~$+$05:05:31.7 & 1.1136 & 18.699            \\
105017.31$+$012450.9 & 10:50:17.31~~$+$01:24:50.9 & 1.2007 & 18.800            \\
105048.25$+$032328.6 & 10:50:48.25~~$+$03:23:28.6 & 1.1509 & 18.914            \\
105118.61$+$015755.9 & 10:51:18.61~~$+$01:57:55.9 & 1.1812 & 18.305            \\
105229.50$+$031131.5 & 10:52:29.50~~$+$03:11:31.5 & 1.1665 & 18.832            \\
105234.24$-$001501.0 & 10:52:34.24~~$-$00:15:01.0 & 1.1607 & 17.694            \\
105352.72$+$050043.9 & 10:53:52.72~~$+$05:00:43.9 & 1.1320 & 18.865            \\
105414.09$-$001803.5 & 10:54:14.09~~$-$00:18:03.5 & 1.1618 & 18.883            \\
105459.34$+$031151.3 & 10:54:59.34~~$+$03:11:51.3 & 1.1819 & 18.582            \\
105527.67$+$002001.5 & 10:55:27.67~~$+$00:20:01.5 & 1.1448 & 18.782            \\
105543.01$+$001001.7 & 10:55:43.01~~$+$00:10:01.7 & 1.1093 & 19.052            \\
105549.72$+$031324.1 & 10:55:49.72~~$+$03:13:24.1 & 1.1348 & 18.618            \\
105703.23$+$040526.8 & 10:57:03.23~~$+$04:05:26.8 & 1.1330 & 18.078            \\ 
105837.95$+$033124.9 & 10:58:37.95~~$+$03:31:24.9 & 1.1215 & 18.857            \\
105943.44$+$024418.2 & 10:59:43.44~~$+$02:44:18.2 & 1.1024 & 18.618            \\
110121.27$+$023333.0 & 11:01:21.27~~$+$02:33:33.0 & 1.0874 & 18.306            \\
\\
\end {tabular}
\label{lqg_U1.11_table}
\end {table*}

\section{Properties of the Clowes \& Campusano LQG from the SDSS DR7QSO quasars}

The Clowes \& Campusano LQG has been detected by three independent methods
\citep{Clowes1991, Newman1998, Newman1999, Williger2002}. The detection in
the DR7QSO database as U1.28 adds a fourth.

\begin{figure*}
\includegraphics{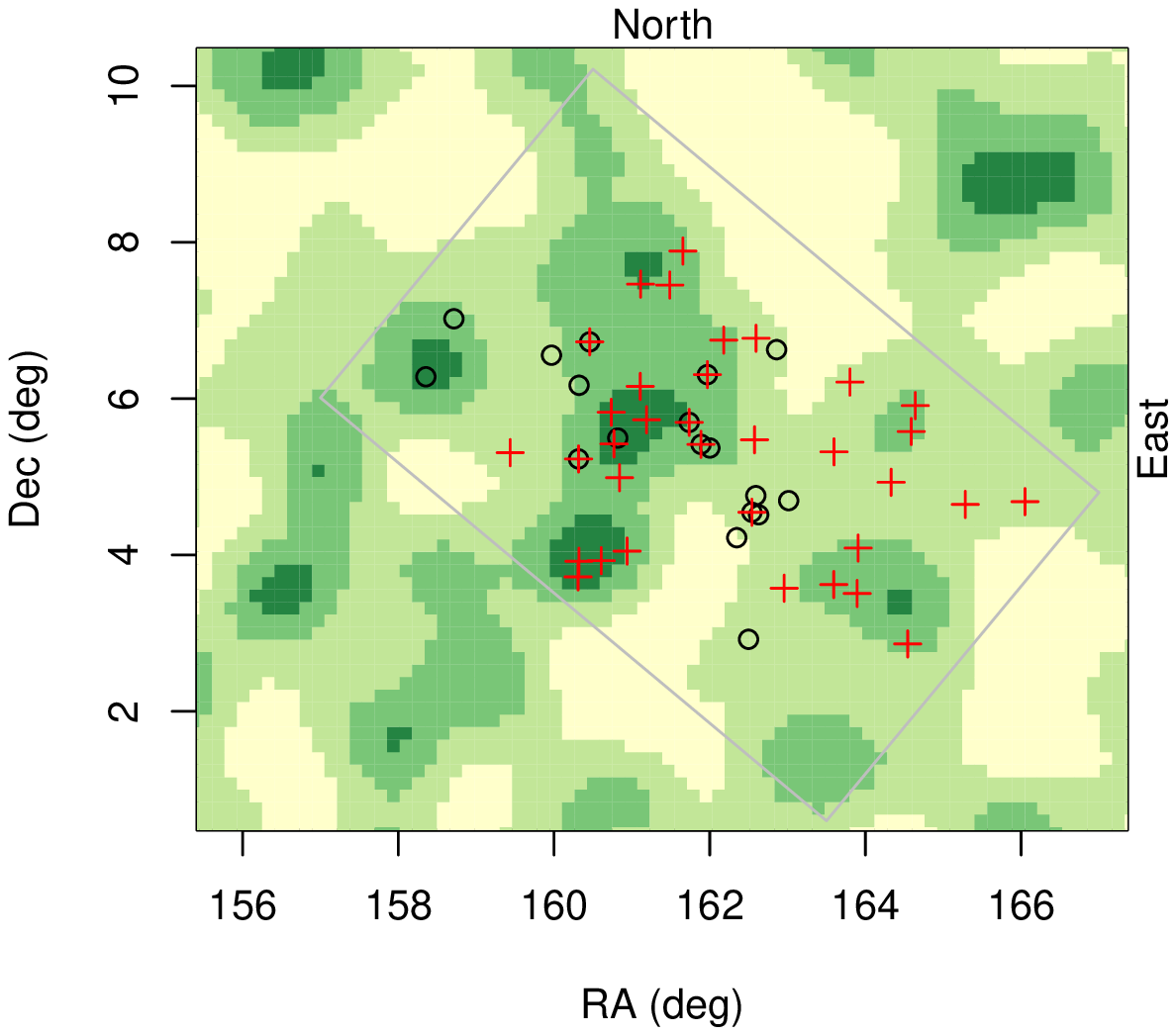}
\caption{The sky distribution of the 34 quasars of U1.28 ($\bar{z} = 1.28$)
(crosses) that are connected at the linkage scale of 100~Mpc together with
the original 18 LQG members (circles). The area shown is approximately
$12^\circ \times 10^\circ$, which is $\sim 6$ times larger than the area
covered by U1.28 itself. The DR7QSO quasars are limited to $i \le
19.1$. Superimposed on these distributions is a kernel-smoothed intensity
map (isotropic Gaussian kernel, $\sigma = 0.5^\circ$), plotted with four
linear palette levels ($\le 0.8$, 0.8--1.6, 1.6--2.4, $\ge 2.4$
deg$^{-2}$), for all of the quasars in the redshift range of U1.28 ($z:
1.1865 \rightarrow 1.4232$). The co-location on the sky of U1.28 and the
original 18 LQG members is clear. Although only the western boundary of the
original LQG members did not encounter the limits of its survey, the
western, northern and southern boundaries of U1.28 and the original 18 are
essentially the same, apart from two compact clumps to the north and
south-west. The eastern boundary of U1.28 extends further than the original
18 by $\sim 2^\circ$. (The grey rectangle shows an area \{A47\} defined for a
following figure.)}
\label{skydist_U1.28}
\end{figure*}

The location on the sky of the 34 members of U1.28 corresponds well to that
of the original LQG: the western, northern and southern boundaries seem to be
essentially the same, apart from two compact clumps to the north and
south-west, (Fig.~\ref{skydist_U1.28}), while the eastern boundary is
extended by $\sim 2^\circ$. Note that for the original 18 only the western
boundary did not encounter the limits of the survey, but the extension seen
with U1.28 is predominantly eastwards, with no major extensions either
northwards or southwards. The coincidence of these sets of 18 and 34 seems
particularly striking given that only six quasars are in common (of ten
possible for $i \le 19.1$).

\begin{figure*}
\includegraphics{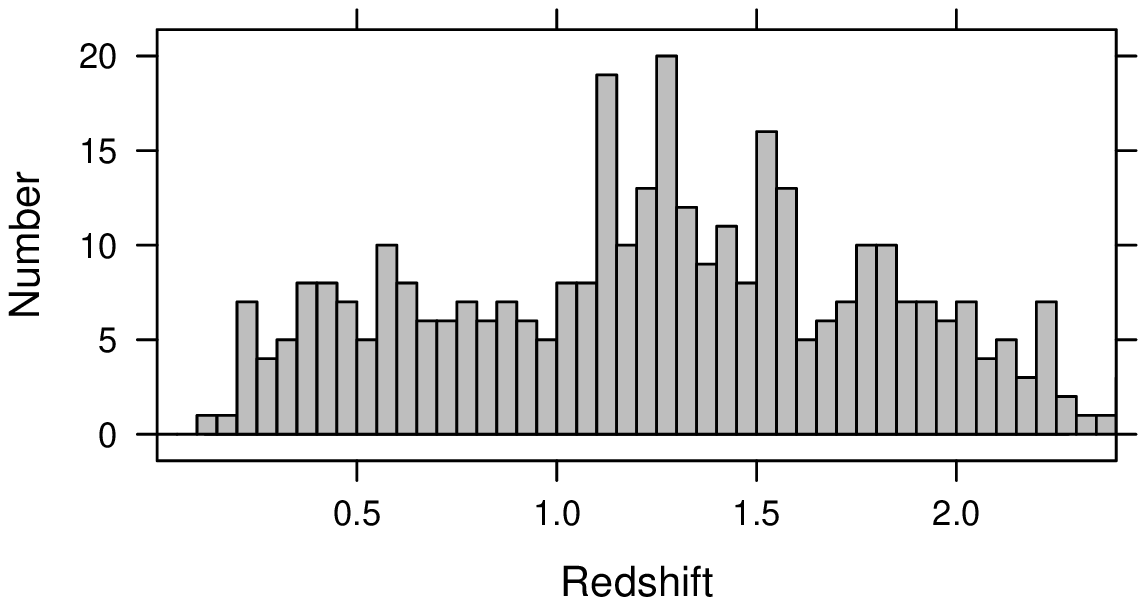}
\includegraphics{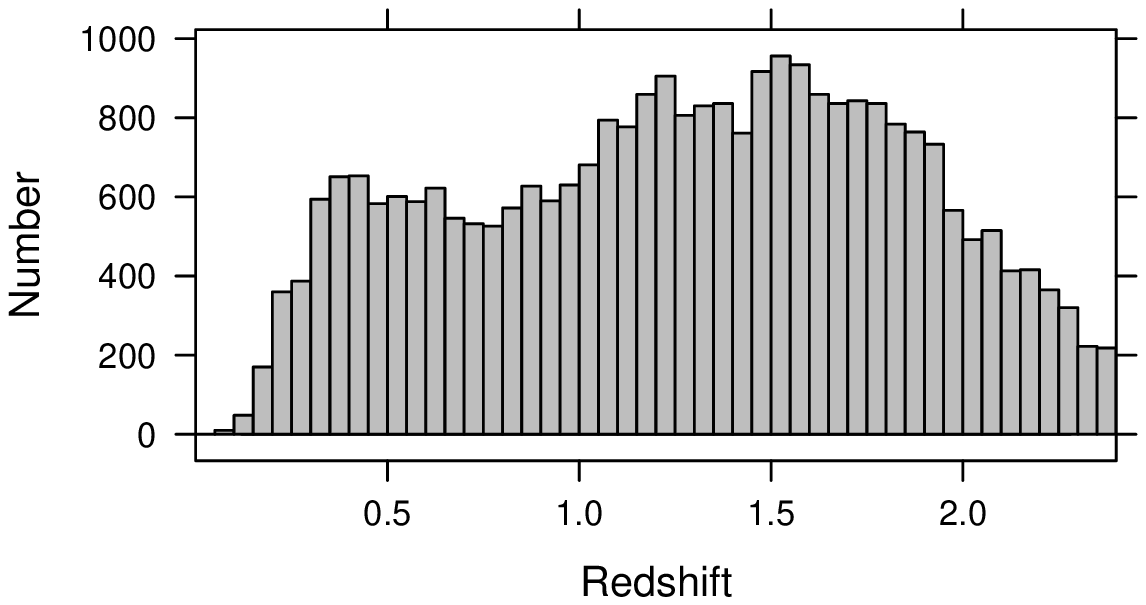}
\caption{The upper histogram shows the redshift histogram for all DR7QSO
quasars ($i \le 19.1$) in the area A47 that contains both U1.28 and the
original members of the LQG. The area A47 is shown by the grey rectangle in
Fig.\ref{skydist_U1.28}. The lower histogram shows the redshift distribution
($i \le 19.1$) of the control area A3725 for comparison. Both histograms
have been restricted to $z \le 2.4$ for clarity. The histogram for A47 shows
a prominent peak for $1.20 < z \le 1.35$, which corresponds to U1.28. Note
also the peaks for $1.10 < z \le 1.15$ and $1.50 < z \le 1.60$.}
\label{zdist_A47}
\end{figure*}

The intensity map of Fig.~\ref{skydist_U1.28} is many times ($\sim 6$) larger
than the area covered by U1.28 itself. In the upper histogram of
Fig.~\ref{zdist_A47} we show the redshift distribution for quasars in a
smaller rectangular area of $\sim 47$ deg$^{-2}$ (A47, actually 46.3 deg$^2$)
that contains both U1.28 and the original members of the LQG. The limits of
this area are shown by the grey rectangle in Fig.~\ref{skydist_U1.28}. In
the lower histogram we show the redshift distribution of the control area
A3725 for comparison. Both of the histograms are derived from the DR7QSO
catalogue restricted to $i \le 19.1$, and for clarity they have been further
restricted to $z \le 2.4$.

The histogram for A47 (upper) shows a prominent peak for $1.20 < z \le 1.35$,
which corresponds to U1.28. Note also the peaks for $1.10 < z \le 1.15$ and
$1.50 < z \le 1.60$, which will be discussed further below.

\begin{figure*}
\includegraphics{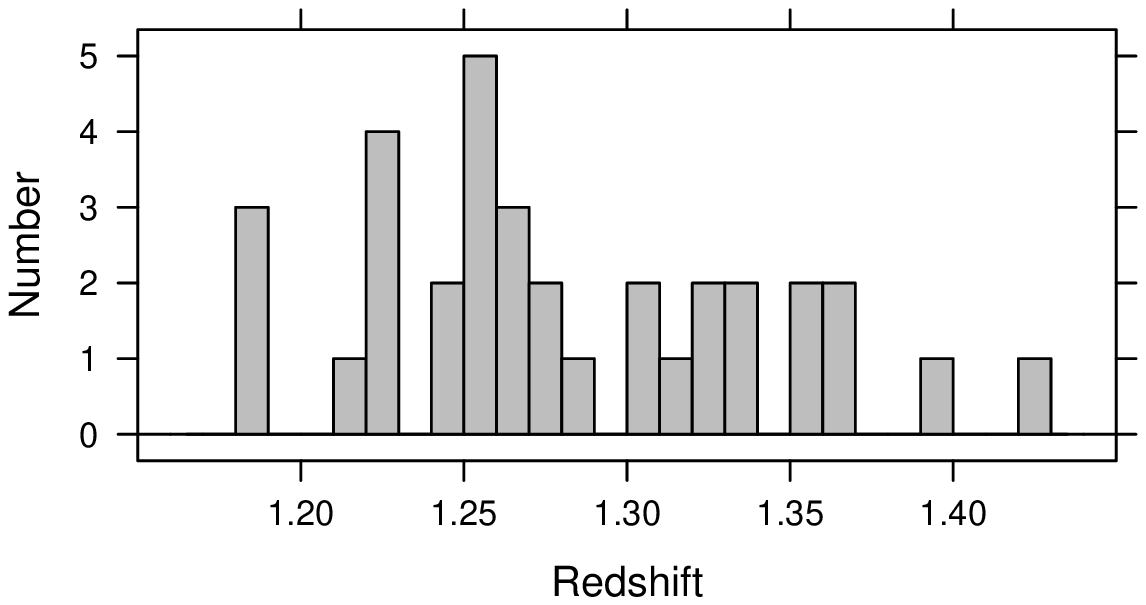}
\caption{The redshift distribution within the unit U1.28 of 34 100Mpc-linked
quasars that corresponds to the Clowes \& Campusano LQG. The histogram is
for a bin size of $\Delta z = 0.01$.}
\label{zdist_U1.28}
\end{figure*}

Fig.~\ref{zdist_U1.28} shows the redshift distribution within the unit U1.28
of 34 100Mpc-linked quasars that corresponds to the Clowes \& Campusano LQG.
Although there are no particularly compelling features in the distribution,
there is possibly some concentration of redshifts to the lower half of the
range.

\section{A new LQG at \lowercase{z}=1.11 in the same direction}

While investigating the appearance of the Clowes \& Campusano LQG in the SDSS
data two further (candidate) LQGs became apparent in the same general
direction on the sky. We were considering only units with a minimum
membership of 20 at the 100~Mpc linkage scale.

These additional candidate LQGs are designated as U1.11 and U1.54, from their
mean redshifts. U1.11 has 38 members, $\bar{z} = 1.11$, and angular
separation (of RA, Dec centroids) of 1.97$^\circ$ from U1.28. U1.54 has 21
members, $\bar{z} = 1.54$, and angular separation of 1.62$^\circ$ from U1.28.
We present below a method for assessing the statistical significance and
overdensity of groups found by linkage of points. We find that U1.28 and
U1.11 are significant, but U1.54 is not. However, we note that U1.54
corresponds to a known LQG (of marginal significance) at $\bar{z} = 1.53$
(or, as published, median $z = 1.51$) that was discovered by
\citet{Newman1998} and \citet{Newman1999} in an independent UV-excess survey
(Chile-UK Quasar Survey). Thirteen members were found in the original
discovery, of which 10 are present in this re-discovery.

The peaks in the upper redshift histogram of Fig.~\ref{zdist_A47} for the
intervals $1.10 < z \le 1.15$ and $1.50 < z \le 1.60$, which were mentioned
briefly in the earlier discussion of that figure, correspond to U1.11 and
U1.54.

\begin{figure*}
\includegraphics{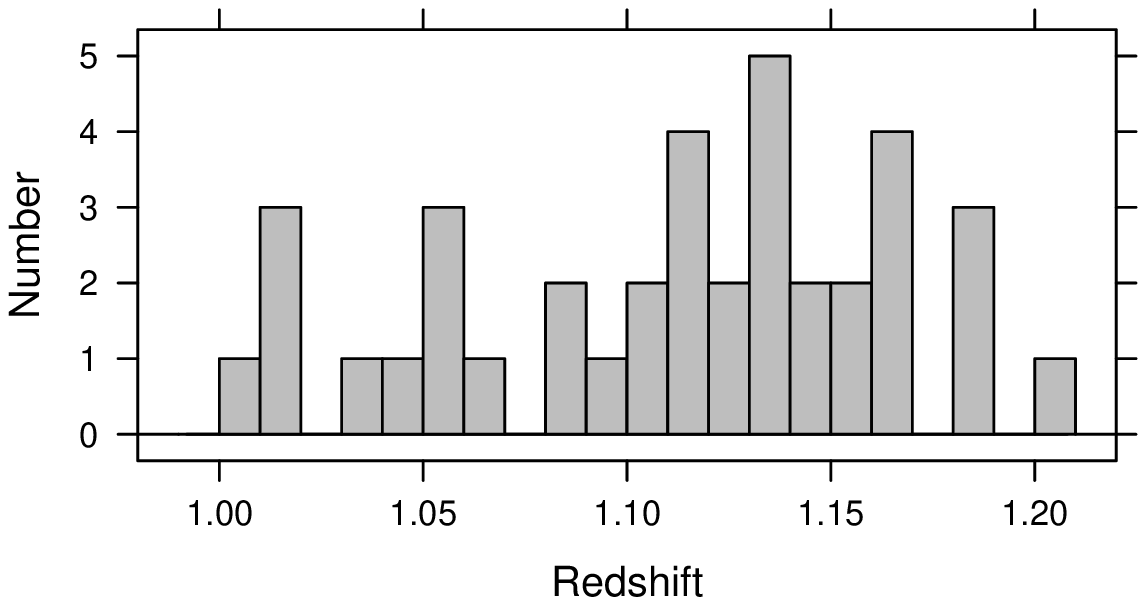}
\caption{The redshift distribution within the unit U1.11 of 38 100Mpc-linked
quasars. The histogram is for a bin size of $\Delta z = 0.01$.}
\label{zdist_U1.11}
\end{figure*}

U1.11, however, is a new discovery, notable for both its appearance in the
same cosmological neighbourhood as U1.28 and for its similarly large number
of members. The 38 quasars of U1.11 are listed in
Table~\ref{lqg_U1.11_table}. The distribution of redshifts for U1.11 is shown
in Fig.~\ref{zdist_U1.11}. Again, there are no particularly compelling
features in the distribution, but in this case there is possibly some
concentration of redshifts to the upper half of the range.

\begin{figure*}
\includegraphics[height=70mm]{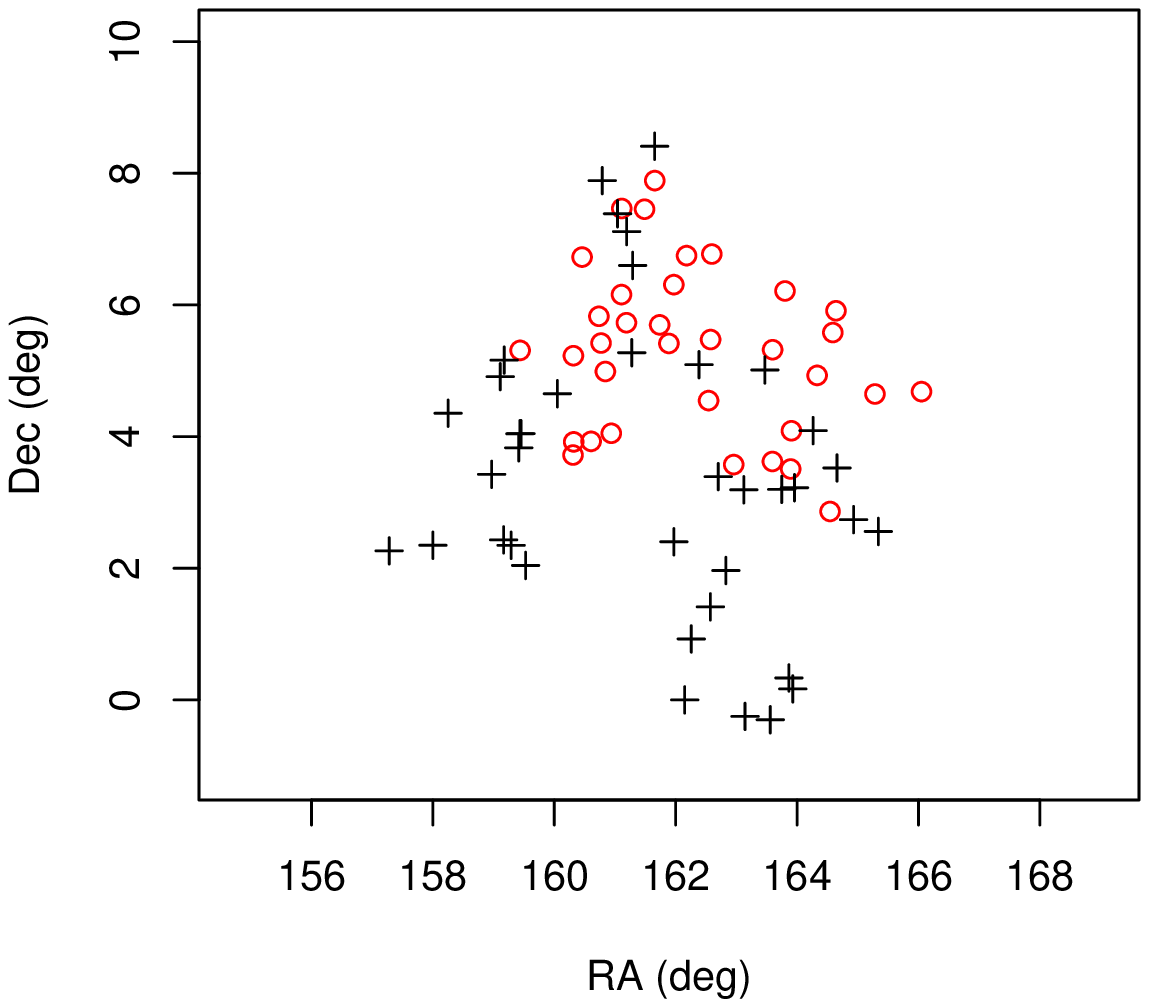}
\includegraphics[height=70mm]{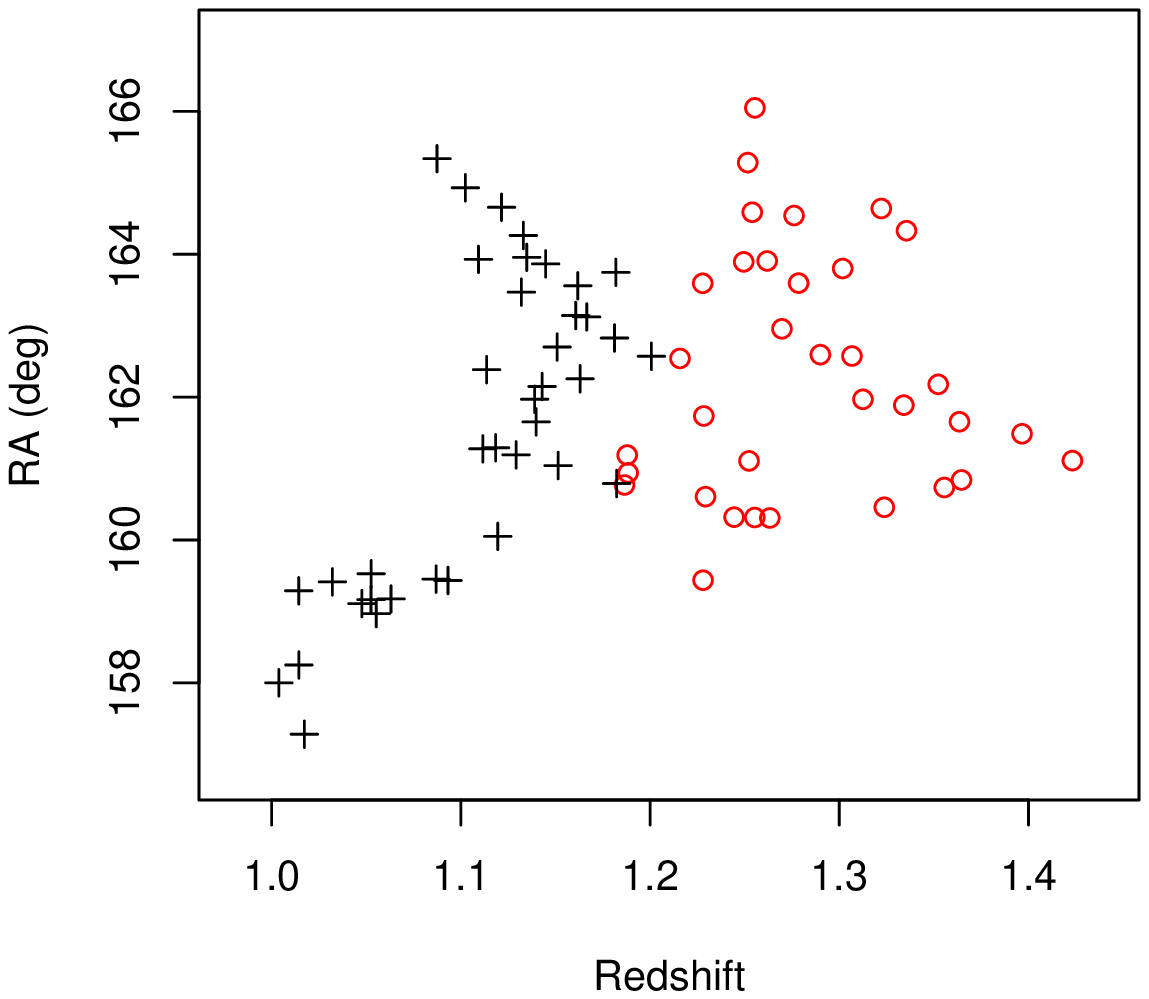}
\includegraphics[height=70mm]{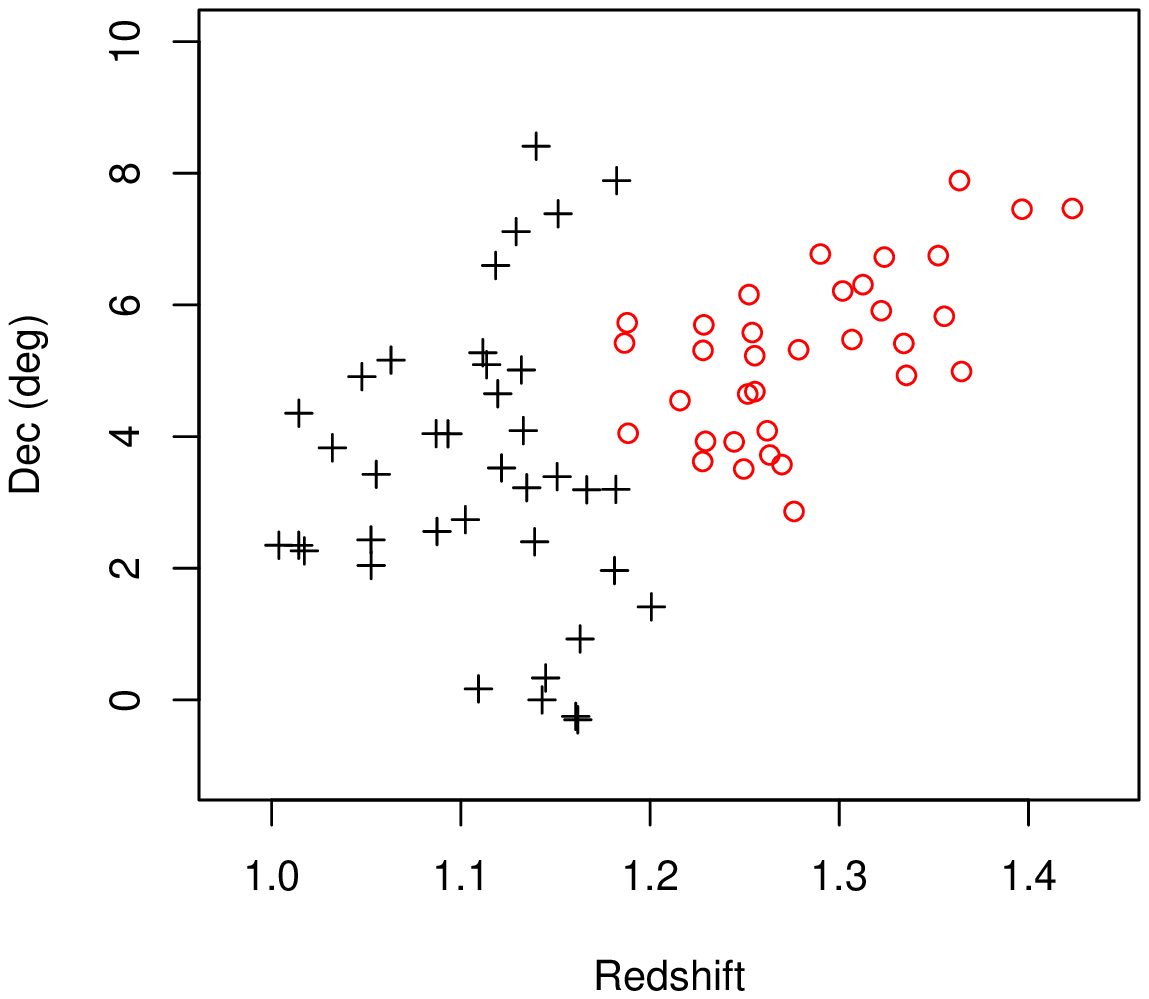}
\caption{Projections of the spatial distributions of the 100Mpc-linked units
of U1.11 (crosses) and U1.28 (circles), shown as plots of: Dec against RA;
RA against redshift; and Dec against redshift.}
\label{units_project}
\end{figure*}

Figure~\ref{units_project} shows three projections (Dec-RA, RA-$z$, Dec-$z$)
of the spatial distributions of U1.11 and U1.28.

In the entire DR7QSO catalogue (i.e.\ area A9380) we find a total of 15 LQG
candidates of such high membership ($N \ge 34$). Of these, U1.28 and U1.11
are the closest, with a separation of centroids of $\sim 410$~Mpc. From the
density of such candidates, the probability of a pair within this separation
occurring by chance somewhere within the coverage of the whole catalogue (and
$1.0 \le z \le 1.8$) is $\sim 0.5$, and so this pair is consistent. Note,
however, that U1.28 and U1.11 do appear to be quite distinct: a small
increase in the linkage scale does not lead to their merger as a single
unit. The volume occupied by U1.28 and U1.11 together thus appears rather
distinctive for quasars on a large scale and would presumably correspond to
some notable features in the cosmic web if the distribution of galaxies was
accessible to observation. As mentioned above, and as is apparent from
Figure~\ref{units_project}, U1.28 and U.11 are also quite closely aligned
with the line of sight.

\section{Assessment of statistical significance and overdensities}

Groups found by the linkage of points generally require a separate procedure
to assess statistical significance. \citet{Graham1995} considered this
problem previously, and created the $m, \sigma$ method, which requires that
groups show significant sub-clustering; \citet{Pilipenko2007} addressed it
differently by counting the frequency of comparable groups in random sets. A
procedure is also needed to estimate overdensities. \citet{Pilipenko2007}
estimated overdensity from properties of the MST, which method is convenient
but was not justified in that paper. In this section we present a new
procedure based on the convex hull that we can use for both statistical
significance and overdensity.

We use a measure of the volume occupied by a LQG to assess the statistical
significance and estimate the overdensity: a LQG must occupy a smaller volume
than the expectation for the same number of random points.

A simple way to estimate the volume is to define a ``RA-Dec-z box'' with
corners determined by the RA, Dec, z limits of the LQG. However, since LQGs
are typically not cuboids aligned with the coordinate axes, the box typically
overestimates the volume quite severely.  The RA-Dec-z boxes will generally
also include non-members and so their volumes are not then solely those of
the LQGs.

Another way to estimate the volume is to use the volume of the convex hull
\footnote{The 3D convex hull of a set of points is the polyhedron of minimum
  volume that contains the lines connecting all pairs of points.}. However,
the convex hull typically underestimates the volume (although overestimates
are also possible, depending on morphology) because it can wrap tightly
around the surface points and does not then consider any surrounding region
as belonging to them.

We can construct a measure of volume that should better reflect the volume
occupied by the LQG than either the RA-Dec-box or the convex hull of the
points. We do this by expanding each member point of a unit to be a sphere,
with radius set to be half of the mean linkage (MST edge length) of the
unit. In this way, each point is associated with a spherical volume. We then
take the volume of the LQG to be the volume of the convex hull of these
spheres. We shall refer to this method as the CHMS method --- convex hull of
member spheres.  Note that this measure refers to the LQG members only,
unlike the RA-Dec-z box, which typically incorporates non-members too.

The distribution of CHMS volumes resulting from random points that have been
distributed with known density allows the statistical significance of a LQG
to be assessed. The CHMS volumes for sets of random points can also be used
to estimate residual biases and, if required, make corrections to the CHMS
volumes, densities and overdensities for LQGs.

The residual bias is expressed as a volume correction, which is the ratio of
the known volume of the sets of random points to their mean CHMS volume. It
corrects for imperfections, of consequence only at the lower memberships, in
reproducing the true volume with the above (natural) choice of sphere
radius. The observed CHMS volume for a LQG can then be corrected by
multiplying by the appropriate volume correction.

For assessing the statistical significance of a LQG of membership $N$, we
compare the departure of its CHMS volume (uncorrected, since there is no
need for the corrections here) from the mean of the distribution of CHMS
volume for 1000 random sets of $N$. Each random set is defined by
distributing $N$ points in a cube, of volume such that the density in the
cube corresponds to the density in A3725 for the redshift limits of the
LQG. In this way, we find that the departures from random expectations for
U1.28, U1.11 and U1.54 are respectively $3.57\sigma$, $2.95\sigma$, and
$1.75\sigma$.

U1.54 thus appears as not significant, although, as we have mentioned above,
it is a re-discovery, with an independent sample, of a LQG that was
previously known \citep{Newman1998, Newman1999}, albeit at marginal
significance then also.

After correcting the CHMS volumes for residual bias the estimated
overdensities of the significant LQGs are $\delta_q = \delta \rho_q / \rho_q
= 0.83, 0.55$ for U1.28 and U1.11 respectively. (The volume corrections are
$\sim$ 10 and 8 per cent respectively.)

\citet{Pilipenko2007} gives a method for determining the overdensity of a LQG
that avoids defining the containing volume, but gave no justification for it
and did not discuss possible biases. In this method, the overdensity is
defined to be $\delta_q = {\langle l_0^3 \rangle} / {\langle l^3 \rangle}$,
where $l$ is MST edge-length for the LQG and $l_0$ is MST edge-length for a
control area elsewhere. Here we modify this definition by including the
``$-1$'' that is more usual for overdensities, giving $\delta_q = ({\langle
  l_0^3 \rangle} / {\langle l^3 \rangle}) - 1$.  Recall that the MST is
equivalent to the single-linkage hierarchical clustering used in this
paper. For this method we obtain $\delta_q = 0.78, 1.31$ for U1.28 and U1.11
respectively, with $l_0$ having been obtained from the control area A3725
separately for the redshift range of each unit. The overdensities by the
Pilipenko method are thus higher for U1.11 than for our CHMS method.

The CHMS overdensities here, $\delta_q \sim$ 0.5-0.9, are substantially lower
than those found by \citet{Pilipenko2007} and a little higher than those
found by \citet{Miller2004}. The two Pilipenko categories, as mentioned above
are: (i) size $\sim$ 85~Mpc, membership $\sim$ 6-8, overdensity $\sim$ 10 (or
9 with the ``$-1$''); and (ii) size $\sim$ 200~Mpc, membership $\ga 15$,
overdensity $\sim$ 4 (or 3 with the ``$-1$''). The overdensities from
\citet{Miller2004} are calculated for spherical top-hat filtering, giving
$\delta_q < 0.44$ for 100$h^{-1}$~Mpc diameters and $\delta_q < 0.17$ for
200$h^{-1}$~Mpc diameters.

\section{Discussion and conclusions}

This paper has re-examined the \citet{Clowes1991} LQG, originally known with
18 members, using data from the SDSS DR7QSO catalogue. It is found as the
unit U1.28 of 34 100Mpc-linked quasars, with mean redshift $\bar{z} = 1.28$.
While the western, northern and southern boundaries remain essentially
unchanged, apart from two compact clumps to the north and south-west, there
is an extension eastwards, beyond the original survey, of $\sim 2^\circ$.

A new LQG, U1.11, was discovered in the same direction --- 1.97$^\circ$ from
U1.28. It has 38 members and mean redshift $\bar{z} = 1.11$. A third
candidate, U1.54, in the same direction at 1.62$^\circ$ from U1.28, appeared
statistically insignificant, although it is a re-discovery of a known
(marginal) LQG \citep{Newman1998, Newman1999}.

We have also presented the ``CHMS method'' for assessing the statistical
significance and overdensity of groups such as LQGs that have been found by
linkage of points.

Attention was first drawn to peculiarities in this area of sky by Cannon \&
Oke ($\sim$ 1980, private communication) who, in the early days of quasar
surveys, noted the unusual ease with which they could find the then
hard-to-find quasars with $z \sim$ 1.0-1.6. It may be that, now we know there
are not one but two LQGs of unusually high membership in this direction, we
finally have the complete explanation of their result.

A simple measure of the characteristic size of the LQGs, which takes no
account of morphology, is the cube root of the corrected CHMS volumes, giving
$\sim 350, 380$~Mpc for U1.28 and U1.11 respectively. Clearly these are very
large sizes, placing these two LQGs among the largest features so far seen in
the early universe. For comparison, \citet{Yadav2010} give an idealised limit
to the scale of homogeneity in the concordance cosmology as $\sim$ 370~Mpc:
it should not be possible to find departures from homogeneity above this
scale. The LQGs appear to be only marginally consistent with this scale of
homogeneity.

Calculation of the inertia tensor for these two LQGs shows a ratio $\sim 2.5$
for the longest and shortest principal axes in both cases. They are therefore
substantially elongated. By this measure the longest axes are $\sim$ 630 and
780~Mpc for U1.28 and U1.11 respectively. Their morphologies appear to be
markedly oblate, like a thick lens, each with two comparably large long axes
and a short axis that is smaller by the factor $\sim 2.5$. Clearly the long
axes do exceed the expected scale of homogeneity.

The estimated overdensities are $\delta_q = \delta \rho_q / \rho_q = 0.83$,
$0.55$ for U1.28 and U1.11 respectively. These overdensities are substantially
lower than those found by \citet{Pilipenko2007} and a little higher than
those found by \citet{Miller2004}.

The occurrence of structure on a particular scale is naturally taken to mean
that the universe is not homogeneous on that scale. These two LQGs, U1.28 and
U1.11, as overdensities of the amplitudes and scales indicated, thus raise a
question of compatibility with the scale of homogeneity in the concordance
cosmology, if the \citet{Yadav2010} fractal calculations are adopted as
reference. A counter argument could be made that these LQGs are chance
associations of groups on sub-homogeneity scales, analogous to the finding of
\citet{Einasto2011} that the component superclusters of the Sloan Great Wall
have different evolutionary histories. Even if this were true, homogeneity
asserts that any global property of sufficiently large volumes should be the
same within the expected statistical variations, so the density of quasars in
these LQGs would remain distinctive. Of course, unknown observational biases
or selection effects could conceivably also affect the overall dimensions of
the LQGs.

In finding compatibility of LQGs with concordance cosmology
\citet{Miller2004} noted that they had not considered questions of shape and
topology. Given both the large sizes and elongated morphology that we find
for U1.28 and U1.11 we have begun a programme to re-investigate
compatibility, using the full set of LQGs that we find from the DR7QSO
catalogue.

\section{Acknowledgments}

LEC received partial support from Center of Excellence in Astrophysics
and Associated Technologies (PFB 06).

The many authors of GNU/Linux, R, GGobi and TOPCAT are gratefully
acknowledged.

This research has used the SDSS DR7QSO catalogue \citep{Schneider2010}.

Funding for the SDSS and SDSS-II has been provided by the Alfred P. Sloan
Foundation, the Participating Institutions, the National Science
Foundation, the U.S. Department of Energy, the National Aeronautics and
Space Administration, the Japanese Monbukagakusho, the Max Planck
Society, and the Higher Education Funding Council for England. The SDSS
Web Site is http://www.sdss.org/.

The SDSS is managed by the Astrophysical Research Consortium for the
Participating Institutions. The Participating Institutions are the American
Museum of Natural History, Astrophysical Institute Potsdam, University of
Basel, University of Cambridge, Case Western Reserve University, University
of Chicago, Drexel University, Fermilab, the Institute for Advanced Study,
the Japan Participation Group, Johns Hopkins University, the Joint Institute
for Nuclear Astrophysics, the Kavli Institute for Particle Astrophysics and
Cosmology, the Korean Scientist Group, the Chinese Academy of Sciences
(LAMOST), Los Alamos National Laboratory, the Max-Planck-Institute for
Astronomy (MPIA), the Max-Planck-Institute for Astrophysics (MPA), New Mexico
State University, Ohio State University, University of Pittsburgh, University
of Portsmouth, Princeton University, the United States Naval Observatory, and
the University of Washington.

The anonymous referee is thanked for helpful comments.

\bsp

\label{lastpage}

\end{document}